\documentclass[showpacs,nofootinbib]{revtex4}

\usepackage{amsmath,amssymb}
\usepackage{graphicx,psfrag,float,epsfig,color}
\usepackage{mathrsfs,wasysym}
\usepackage{epstopdf}

\DeclareSymbolFontAlphabet{\mathrsfs}{rsfs}
\DeclareMathAlphabet{\mathcal}{OMS}{cmsy}{m}{n}

\newcommand{\scri}{\mathrsfs{I}}
\newcommand{\be}{\begin{equation}}
\newcommand{\ee}{\end{equation}}

\def\tg{{\tilde{g}}}
\def\rhoo{\rho_{\mathrm{ext}}}
\def\rhoi{\rho_{\mathrm{int}}}
\def\Hext{H_{\mathrm{ext}}}

\def\Omext{\Omega_{\mathrm{ext}}}
\def\Omint{\Omega_{\mathrm{int}}}
\def\tphi{\tilde{\phi}}
\def\tA{\tilde{A}}
\def\tB{\tilde{B}}
\def\tC{\tilde{C}}

\begin{document}


\title{Spacelike matching to null infinity}

\author{An{\i}l Zengino\u{g}lu, Manuel Tiglio} 

\affiliation{Department of Physics, and Center for Scientific
  Computation and Mathematical Modeling, University of Maryland,
  College Park, MD 20742, USA}

\begin{abstract}
We present two methods to include the asymptotic domain of a
background spacetime in null directions for numerical solutions of
evolution equations so that both the radiation extraction problem and
the outer boundary problem are solved. The first method is based on
the geometric conformal approach, the second is a coordinate based
approach. We apply these methods to the case of a massless scalar wave
equation on a Kerr spacetime. Our methods are designed to allow
existing codes to reach the radiative zone by including future null
infinity in the computational domain with relatively minor
modifications. We demonstrate the flexibility of the methods by
considering both Boyer-Lindquist and ingoing Kerr coordinates near the
black hole. We also confirm numerically predictions concerning tail
decay rates for scalar fields at null infinity in Kerr spacetime due
to Hod for the first time.
\end{abstract}


\maketitle
\section{Introduction}

In numerical calculations of radiative fields, it is common to
artificially truncate the computational domain by introducing an outer
boundary into the spacetime. It is well known that this practice
introduces conceptual and operational problems, specifically the outer
boundary and the radiation extraction problems.
\cite{Abrahams:1997ut, Sarbach04, Rinne06, Rinne:2007ui, Pazos06,
  Lehner:2007ip, Buchman07, Kreiss:2008ig, Friedrich99,
  Friedrich:2009tq, Seiler:2008hm, Deadman:2009ds, Gallo:2008sk,
  Nerozzi:2008ng, Boyle:2007ft, Boyle:2009vi, Lindblom:2008cm}.  Each
incremental development has taken considerably more effort, while a
clean solution to the problems is available on a geometrical level,
namely including the asymptotic domain on the computational grid
\cite{Penrose63, Penrose65, Tamburino:1966zz, Friedrich83a, Huebner96,
  Husa01, Frauendiener04}.

First attempts on including the asymptotic domain on the numerical
level have been based on using a compactifying coordinate along
outgoing null surfaces \cite{Isaacson:1983xc, Winicour:2008tr}. This
allows one to include null infinity in the computational domain where
no boundary conditions are required and radiation is naturally and
unambiguously measured by idealized observers. In addition, the
numerical evolution is very efficient as the grid follows outgoing
radiation. Setting the numerical outer boundary to null infinity
solves the aforementioned problems in a numerically efficient manner.

Unfortunately null foliations can be inconvenient, especially because
they tend to develop caustics in dynamical spacetimes
\cite{Friedrich83}. For solving the outer boundary and radiation
extraction problems, however, we only need access to null infinity,
whereas the interior foliation can be chosen arbitrarily. This
observation suggests that one could match an arbitrary spacelike slice
in the interior to an asymptotic domain that extends to null
infinity. This has first been tried using Cauchy-characteristic
matching \cite{Anderson88, Bishop93,
  Bishop-etal-1996:Cauchy-characteristic-matching,
  Winicour:2008tr}. There, one uses an evolution based on Cauchy-type
foliations in the interior and communicates the in- and outgoing
characteristic variables along a timelike surface with a
characteristic evolution in the asymptotic domain. This method
introduces numerical difficulties due to the presence of a nonsmooth
matching along which the causal nature of the slicing changes from
spacelike to null. While Cauchy-characteristic matching is a promising
approach that requires further study, in this paper we will explore
an alternative approach to the problem.

Instead of matching spacelike slices to null slices along a timelike
surface, here we use everywhere spacelike slices that approach null
infinity in the asymptotic domain. The latter are called hyperboloidal
surfaces as their behavior is similar to the standard hyperboloids in
Minkowski spacetime \cite{Friedrich83a}. They are more flexible than
null surfaces because the only local condition restricting their
choice is that they are spacelike. The asymptotic condition that they
approach null infinity does not restrict the type of spacelike
surfaces in the interior.  It is favorable for numerical applications
within the hyperboloidal approach that a gauge is chosen in which the
location of future null infinity (scri) is independent of time,
i.e.~scri fixing \cite{Frauendiener98b, Andersson:2002gn, Husa:2005ns,
  Misner:2005yz, Zenginoglu:2007jw, Moncrief:2008ie}. Numerical
experiments with scri fixing have already been made successfully in
spherically symmetric spacetimes \cite{Fodor04,Fodor:2006ue,
  vanMeter:2006mv, Zenginoglu:2008uc, Zenginoglu:2008wc, bizon-2008}.

In this paper we discuss the case of a fixed Kerr black hole
background from the point of view of the hyperboloidal approach with
scri fixing. One might think that the characteristic approach should
be sufficient for calculating radiative fields at null infinity when a
background has been given. Indeed, for given Minkowski and
Schwarzschild backgrounds, null coordinates are very useful and have
been regularly applied in numerical calculations
\cite{Winicour:2008tr, Campanelli:2000in, Gundlach:1993tp,
  Husa:2001pk}. In Kerr spacetime, however, they are difficult to deal
with. There has been an ongoing effort to find a metric representing
the Kerr geometry in Bondi-Sachs form that can be used in numerical
computations \cite{Venter:2005cs, FletcherLun, Bai:2007rs,
  Pretorius:1998sf, Winicour:2008tr}, but to our knowledge no
numerical calculations including null infinity could be presented so
far.

The difficulties with the characteristic approach in Kerr spacetime
arise, in part, from the fact that null coordinates are very rigid,
i.e.~the coordinate freedom for choosing a null surface is more
restricted than for a spacelike surface. The flexibility of
hyperboloidal surfaces, on the other hand, allows us to evolve
radiative fields on a Kerr spacetime up to and including null infinity
while keeping, for example, standard Boyer-Lindquist or ingoing Kerr
coordinates near the black hole \footnote{We use the naming convention
  for the coordinatization of Kerr spacetime as in
  \cite{Burko:2007ju}.}. We have chosen these foliations as they are
the most common ones and have a differing qualitative behavior near
the horizon. In principle, any foliation near the black hole can be
chosen. The construction of such hyperboloidal surfaces in Kerr
spacetime is relatively new \cite{Zenginoglu:2007jw}. In this paper
we show that it is also amenable to numerical calculations.

We discuss two methods for the numerical implementation of the
hyperboloidal matching idea on the example of a scalar wave
equation. In both methods matching includes certain choices of
rescaling, time transformation and coordinate compactification. The
first method is the well-known conformal method as introduced by
Penrose \cite{Penrose63, Penrose65}. It requires knowledge
of the conformal transformation behavior of the equation one is trying
to solve. The second method is an essentially equivalent method that
may be more appealing to researchers unfamiliar with conformal
techniques. It considers the equation to be given in a certain
coordinate system and applies the above-mentioned transformations
\cite{Zenginoglu:2008uc}. These methods allow us to solve the
equations of interest including null infinity and can be applied
within the hyperboloidal approach independent of matching.

We test our implementations by reproducing previously obtained
quasinormal mode frequencies \cite{Krivan97a, Dorband:2006gg} and
tail decay rates \cite{Burko:2007ju, Gleiser:2007ti, Tiglio:2007jp}.
Our numerical tests focus on scalar fields but our method of including
null infinity can be used with a wide variety of systems of evolution
equations on a Kerr, Schwarzschild or Minkowski background that admit
asymptotically flat solutions. In particular, the application of our
techniques to calculate other types of radiative fields such as
gravitational perturbations should be straightforward.

The paper is organized as follows. In Sec.~II we present the main
idea of hyperboloidal matching that allows us to include null infinity
in the computational domain while using standard coordinates near the
black hole. Sec.~III presents the conformal method and applies it
using ingoing Kerr coordinates, Sec.~IV presents the coordinate
based method applied to Boyer-Lindquist coordinates. The resulting
equations are then solved in Sec.~V using standard numerical
techniques. Beyond confirming well-known predictions with our code, we
also study decay rates for scalar perturbations of a Kerr black hole
near and at null infinity confirming predictions due to Hod
\cite{Hod:1999rx} for the first time. We conclude with a short
discussion of our results.

\section{The main idea}\label{sec:main}

As mentioned in the introduction, the construction of null coordinates
in a Kerr background is relatively complicated \cite{Venter:2005cs,
  FletcherLun, Pretorius:1998sf, Bai:2007rs} whereas hyperboloidal
foliations that are useful for numerical purposes can be given
explicitly \cite{Zenginoglu:2007jw}. Following
\cite{Zenginoglu:2007jw}, we construct hyperboloidal foliations that
coincide in an interior domain exactly with ingoing Kerr or
Boyer-Lindquist coordinates.

The ingoing Kerr and Boyer-Lindquist representations of the Kerr
metric will be given in terms of coordinates that we denote by
$(t,r,y,\varphi)$. Here, $t$ is the time coordinate of the
corresponding metric, $r$ is a radial coordinate along level sets of
$t$ and $y$ is an angular coordinate defined by $\cos\vartheta$ where
$\vartheta$ and $\varphi$ are the angular coordinates on a sphere. In
order to match a spacelike surface to null infinity in a stationary
spacetime, we introduce a new time coordinate $\tau$ of the form
\be \label{eq:trafo} \tau = t-h(r).\ee The function $h(r)$ is called
the height function. The advantage of the above transformation is that
it keeps the time direction invariant.  The stationary Killing field
that is timelike outside the event horizon has the same coordinate
representation in both $t$ and $\tau$ for any choice of $h(r)$. It is
given by $\partial_t$ or, equivalently, by $\partial_\tau$. This
implies that the Kerr metric in the new coordinates is manifestly
stationary and natural observers that follow integral curves of the
stationary Killing field are again given by surfaces of constant
spatial coordinates. A natural consequence of the choice
(\ref{eq:trafo}) is that null infinity is at a fixed spatial
coordinate location with respect to a time-independent compactifying
coordinate along level sets of $\tau$. This considerably simplifies
the numerical implementation as well as the analysis of the solution
\cite{Husa:2005ns, Zenginoglu:2008wc}.

A compactifying coordinate can be introduced by \be \label{eq:compr} r
= \frac{\rho}{\Omega}, \qquad \textrm{with}\quad \Omega=
\Omega(\rho). \ee This choice implies that the zero set of $\Omega$
corresponds to infinity in terms of the physical coordinate $r$. The
choice of the conformal factor can be made in such a way as to ensure
that future null infinity (denoted by $\scri^+$) is at the coordinate
location $\rho=S$, for example, by setting
$\Omega_{\mathrm{ext}}(\rho) = 1-\rho/S$. In an interior domain near
the black hole we want to use the standard physical coordinates, which
implies $\Omega_{\mathrm{int}}(\rho)=1$. There is a transition between
the interior and the exterior domain where we match the two
functions. Many choices are possible on the transition domain. Here we
set \be \label{eq:match_om}\Omega(\rho) = \left\{
\begin{array}{ll} \Omint(\rho) 
  & \mathrm{for} \quad \rho\leq \rhoi, \\ \Omint(\rho)\,
  e^{-(\rho-\rhoi)^2/(\rho-\rhoo)^2} +
  \Omext(\rho)\,\left(1-e^{-(\rho-\rhoi)^2/(\rho-\rhoo)^2}\right) \qquad &
  \mathrm{for} \quad \rhoi<\rho<\rhoo, \\ \Omext(\rho) & \mathrm{for}
  \quad \rho\geq \rhoo \,.
\end{array}\right.\ee 

Hyperboloidal foliations that coincide with standard Boyer-Lindquist
or ingoing Kerr foliations in the interior can be constructed by
suitably choosing the derivative of the height function $dh/dr =:
H$. Here we choose
\be \label{eq:match_H} H(\rho) = \left\{
\begin{array}{ll} 0
  & \mathrm{for} \quad \rho\leq \rhoi,  \\ 
   \Hext(\rho)\,\left(1-e^{-(\rho-\rhoi)^2/(\rho-\rhoo)^2}\right) \qquad & 
   \mathrm{for} \quad \rhoi<\rho<\rhoo,  \\ 
   \Hext(\rho) & \mathrm{for} \quad \rho\geq \rhoo.
\end{array}\right.\ee 
The choice $H=0$ along with $\Omega=1$ near the black hole ensures
that the foliation and the coordinates are not changed. The function
$\Hext(\rho)$ is different for Boyer-Lindquist and ingoing Kerr
foliations \cite{Zenginoglu:2007jw}. Its choice will be discussed in
later sections.

\section{The conformal method} \label{sec:metric}
The conformal method to study the asymptotic behavior of radiative
fields was introduced by Penrose \cite{Penrose63, Penrose64,
  Penrose65}. The example of the scalar wave equation was the first
system where the hyperboloidal initial value problem was studied
\cite{Geroch77, Wald84}. In this section we will present the conformal
method for the scalar wave equation on a Kerr spacetime.
\subsection{Conformal transformation of the scalar wave equation}
Our notation follows \cite{Zenginoglu:2008wc}, where the conformal
method was applied to calculate tail decay rates for scalar and
Yang-Mills fields at null infinity in a Schwarzschild spacetime. 

Here we are interested in solutions to the scalar wave equation
$\widetilde{\Box}\tilde{\phi}=0$, for a scalar field $\tilde{\phi}$ in
a Kerr spacetime with metric $\tg$. The coordinate transformations
(\ref{eq:trafo}) and (\ref{eq:compr}) result in a representation of
the Kerr metric $\tg$ that is singular at $\Omega=0$. To include null
infinity in a regular way we need to rescale the Kerr metric $\tg$
with $\Omega^2$. The Kerr spacetime is weakly asymptotically simple,
implying that with a suitable choice of the transformations
(\ref{eq:trafo}) and (\ref{eq:compr}) the rescaled metric
$g=\Omega^2\tg$ is smoothly extendable through null infinity where we
have $\{\Omega=0,\ d\Omega\ne 0\}$. The following conformal
transformation rule holds \be\label{eq:conf_meth}
\left(\Box -\frac{1}{6} R \right) \phi = \Omega^{-3}
\left(\tilde{\Box} - \frac{1}{6}\tilde{R}\right) \tilde{\phi}, \quad
\mathrm{with} \quad \phi = \frac{\tilde{\phi}}{\Omega}, \ee where $R$
and $\tilde{R}$ are the Ricci scalars of the rescaled and the physical
metrics $g$ and $\tg$ respectively. The scalar wave equation with
respect to the rescaled metric then becomes \be\label{eq:solve}
\Box\phi -\frac{1}{6} R \phi = 0.\ee Note that the above equation is
invariant under both coordinate and conformal transformations.

\subsection{First order reduction}
The wave equation (\ref{eq:solve}) can be written in any coordinate
system $\{x^\mu\}$ as
\[ g^{\mu\nu}\partial_\mu\partial_\nu \phi = \Gamma^\mu\partial_\mu\phi + 
\frac{1}{6}R\phi, \qquad \mathrm{where} \qquad \Gamma^\lambda :=
g^{\mu\nu}\Gamma_{\mu\nu}^\lambda. \] To bring it to first order form
we introduce the auxiliary variables
$\phi_\mu:=\partial_\mu\phi$. Assuming that the coordinate $x^0$ denotes
the time direction we get the following system of evolution
equations:
\begin{eqnarray}\label{eq:num_ev}
\left(-g^{00}\right)\,\partial_0\phi_0 &=& 2g^{0a}\partial_a\phi_0 +
g^{ab}\partial_a\phi_b - \left(\Gamma^0\phi_0 + \Gamma^a\phi_a +
\frac{1}{6}\,R\,\phi\right), \nonumber \\ \partial_0\phi_a &=&
\partial_a \phi_0, \\ \partial_0 \phi \ &=& \phi_0. \nonumber
\end{eqnarray}
The above system is symmetric hyperbolic \cite{Friedrich:2000qv}. Its
coefficients are calculated with respect to the conformal metric $g$
written in coordinates $\{\tau,\rho,y,\varphi\}$. In the case of a
Kerr spacetime we can apply the separation of variables $\phi(\tau,
\rho, y, \varphi)=\phi(\tau, \rho, y) e^{-i\,k\,\varphi}$ and study
solutions for each $k$ mode independently, because azimuthal modes
decouple in the presence of axisymmetry.  Here, we will choose initial
data corresponding to a single azimuthal mode with $k=0$. In this
case, the indices $a$ and $b$ span over $\rho$ and $y$.
\subsection{Hyperboloidal compactification in ingoing Kerr coordinates}
\label{sec:iK}

In the following, we present a hyperboloidal compactification of the
ingoing Kerr representation of the Kerr metric. The ingoing Kerr
slicing is one of the most commonly used slicing conditions for
numerical calculations on Kerr spacetime. Its level sets allow us to
apply an excision technique inside the black hole
horizon. Asymptotically they approach spatial infinity. For their
relation to the Boyer-Lindquist coordinates see \cite{Burko:2007ju}.
Following usual notation we introduce \be\label{eq:kerrdef}
\widetilde{\Sigma} := r^2 + a^2 y^2, \qquad \widetilde{\triangle} :=
r^2 + a^2 - 2mr. \ee where $m$ and $a$ are the ADM mass and the
angular momentum of the Kerr spacetime, respectively. After
transforming to compactifying coordinates we will use $\Sigma
:=\Omega^2 \widetilde{\Sigma}$ and $\triangle
:=\Omega^2\widetilde{\triangle}$.

The Kerr metric in ingoing Kerr coordinates reads
\begin{eqnarray} 
\label{eq:st_ks} \tilde{g}_{\textrm{iK}} &=&
  -\left(1-\frac{2mr}{\widetilde{\Sigma}}\right) dt^2
  +\frac{4mr}{\widetilde{\Sigma}}\,dt dr - \frac{4 a m
    r}{\widetilde{\Sigma}}(1-y^2) \,dt\,d\varphi -
  2a(1-y^2)\left(1+\frac{2mr}{\widetilde{\Sigma}}\right) \,dr
  d\varphi+ \nonumber \\ && +
  \left(1+\frac{2mr}{\widetilde{\Sigma}}\right)\,dr^2 +
  \frac{\widetilde{\Sigma}}{1-y^2}\,dy^2 + \left( r^2+a^2+
  \frac{2ma^2r (1-y^2)}{\widetilde{\Sigma}} \
  \right)(1-y^2)\,d\varphi^2 \, .
\end{eqnarray}

This metric gives the following simple evolution equation for
$\partial_t \tilde{\phi}_t$: \be\label{eq:stkseq}
(\widetilde{\Sigma}+2mr)\,\partial_t\tphi_t = 4mr\, \partial_r\tphi_t+
\widetilde{\triangle}\,\partial_r\tphi_r+(1-y^2)\,\partial_y\tphi_y+2m\,
\tphi_t+2(r-m)\,\tphi_r-2y\,\tphi_y.\ee

Applying the coordinate transformations (\ref{eq:trafo}) and
(\ref{eq:compr}) to the above metric and rescaling with $\Omega^2$
gives
\begin{eqnarray} \label{eq:kerr_ks}
&g_{\textrm{iK}}& =\ -\Omega^2
    \left(1-\frac{2m\rho\,\Omega}{\Sigma}\right) \,d\tau^2 - 2L
    \,\left( H -
    \frac{2m\rho\,\Omega}{\Sigma}(1+H)\right)\,d\tau\,d\rho - \frac{4
      a m \rho\, \Omega^3}{\Sigma} (1-y^2)\,d\tau\,d\varphi+
    g_{\rho\rho}\, d\rho^2 - \nonumber \\ &-& 2 L
    a(1-y^2)\left(1+\frac{2m\rho\Omega}{\Sigma}(1+H)\right) d\rho
    d\varphi + \frac{\Sigma}{1-y^2} dy^2+ \left( \rho^2+a^2\Omega^2+
    \frac{2ma^2\rho \Omega^3 (1-y^2)}{\Sigma} \right) (1-y^2)
    d\varphi^2.
\end{eqnarray}
Here, $L:=\Omega-\rho\,\partial_\rho\Omega$, and 
\[ g_{\rho\rho} = \frac{L^2}{\Omega^2}\left(1 - H^2+\frac{2m\rho\,\Omega}
{\Sigma}(1+H)^2\right). \]
The radial component of the spacetime metric appears to be singular,
but with a suitable choice for the derivative of the height function
it acquires an explicitly regular form. Following
\cite{Zenginoglu:2007jw} we make that choice in the exterior domain to
be \be
\label{eq:height_ks} \Hext = 1+\frac{4m\Omega}{\rho} +
\frac{(8m^2-C^2_{\textrm{iK}})\Omega^2}{\rho^2}, \ee where
$C_{\textrm{iK}}$ is a free parameter. Only the first two terms are
required for the regularity of $g_{\rho\rho}$. The third term
(involving the free parameter $C_\textrm{iK}$) is taken for numerical
purposes as will be discussed later. The appearance of such a free
parameter for hyperboloidal foliations should be contrasted with
asymptotic solutions to eikonal equations used for asymptotic
constructions of null hypersurfaces, where no such parameters are
allowed \cite{Bai:2007rs}.

With the above choice for the derivative of the height function the
metric is manifestly regular at $\scri ^+$. By setting $\Omext =
1-\rho/S$ we get
\[ g_{\textrm{iK}}|_{\scri^+} = -2\,d\tau\,d\rho+2\,\frac{C^2_{\textrm{iK}}}
{S^2}\,d\rho^2 - 2a (1-y^2)\, d\rho\, d\varphi + \frac{1}{1-y^2}\,dy^2
+ (1-y^2)\,d\varphi^2. \] The coordinate speed of outgoing
characteristics at null infinity reads
$c_+=-2g_{\tau\rho}/g_{\rho\rho}$; for the above metric this is
$c_+=S^2/C^2_{\textrm{iK}}$. We see that the coordinate speed of
characteristics depends on the coordinate location of null infinity
and an additional free parameter $C_{\textrm{iK}}$. The latter is
related to the asymptotic value of the mean extrinsic curvature of our
surfaces. This is the reason why we included the third term in the
expansion (\ref{eq:height_ks}): the parameter $C_{\textrm{iK}}$ is
important if we want the coordinate location of $\scri^+$ to be large
while avoiding strong restrictions on the allowed time step due to a
large value of the coordinate speed of outgoing characteristics.

\section{The coordinate based approach}

We can recognize essentially three steps in the hyperboloidal
compactification of Kerr spacetime: a time transformation
(\ref{eq:trafo}), a spatial compactification (\ref{eq:compr}), and a
rescaling (\ref{eq:conf_meth}). In the previous section we calculated
a conformal metric that has a smooth extension through null infinity
and studied the conformal transformation behavior of the underlying
covariant differential equation. In this section we will instead go
through the above three steps at the level of the partial differential
equations written in some coordinate system. We will follow this
approach in an explicitly noncovariant notation. The rationale behind
this is to have an alternative to the geometric approach, which might
be more straightforward to apply for those not familiar with conformal
techniques. This approach may also be useful in cases where the
conformal transformation behavior of the equation of interest is
difficult to calculate \cite{Zenginoglu:2008uc}.

\subsection{The transformations}

Suppose that we are given a linear, second order partial differential
equation for a function $\tphi$ in two spatial dimensions
\be \label{eq:coord} \partial_t \tphi_t = \tA^{tr}\partial_r\tphi_t +
\tA^{rr}\partial_r \tphi_r + \tA^{yy} \partial_y \tphi_y +
\tB^t\tphi_t + \tB^r\tphi_r + \tB^y\tphi_y + \tC \tphi, \ee where
$\tphi_\mu:=\partial_\mu \tphi$.  Assuming that the coefficients are
independent of $t$, we write the above equation under the following
transformations:
\begin{enumerate}
\item Time transformation: We introduce a new time coordinate as in
  (\ref{eq:trafo}). Equation (\ref{eq:coord}) then becomes
  \begin{eqnarray} \label{eq:hypal} (1-\tA^{rr}H^2+\tA^{tr}H) \, 
\partial_\tau \tphi_\tau &=& (\tA^{tr}-2H\tA^r) \partial_r\tphi_\tau +
\tA^{rr}\partial_r \tphi_r + \tA^{yy} \partial_y \tphi_y +
\nonumber\\ &+& (\tB^t-\tB^r H-\tA^{rr}\partial_r H) \tphi_\tau +
\tB^r\tphi_r + \tB^y\tphi_y + \tC \tphi, \end{eqnarray}
\item Spatial compactification: We introduce a compactifying radial
  coordinate as in (\ref{eq:compr}). We set 
  \[ J:= \frac{d\rho}{dr} = \frac{\Omega^2}{L}, \qquad 
J' := \frac{dJ}{dr} = \frac{\Omega^3}{L^3}(2(\partial_\rho\Omega) L +
\rho\,\Omega\partial_\rho^2\Omega), \] with
$L=\Omega-\rho\,\partial_\rho\Omega$ as before. Equation
(\ref{eq:coord}) becomes \be \label{eq:compify} \partial_t \tphi_t =
\tA^{tr}J\,\partial_\rho\tphi_t + \tA^{rr}J^2\,\partial_\rho
\tphi_\rho + \tA^{yy} \partial_y \tphi_y + \tB^t\tphi_t + (\tB^r J+
\tA^{rr}J')\,\tphi_\rho + \tB^y\tphi_y + \tC \tphi. \ee 
\item Rescaling: We define the rescaled evolution variable $\phi:=f(r)
  \tphi$. Equation (\ref{eq:coord}) becomes
  \begin{eqnarray} \label{eq:resc} 
    \partial_t \phi_t &=& \tA^{tr}\partial_r\phi_t + \tA^{rr}\partial_r
    \phi_r + \tA^{yy} \partial_y \phi_y + \left(\tB^t-\tA^{tr}
    \frac{\partial_r f}{f}\right) \phi_t + \left( \tB^r-2\tA^{rr}
    \frac{\partial_r f}{f}\right) \phi_r + \tB^y\phi_y + \nonumber\\
    &+& \left(\tC-\tB^r\frac{\partial_r f}{f}+ \tA^{rr}
    \left(2\frac{(\partial_rf)^2}{f^2} - \frac{\partial_r^2
      f}{f}\right)\right) \phi \,. \end{eqnarray}
\end{enumerate}
Combining the three steps above gives us finally an evolution
equation of the form \be \label{eq:final} \partial_\tau \phi_\tau =
A^{\tau\rho}\partial_\rho\phi_\tau + A^{\rho\rho}\partial_\rho
\phi_\rho + A^{yy} \partial_y \phi_y + B^\tau\phi_\tau +
B^\rho\phi_\rho + B^y\phi_y + C \phi. \ee 

This method is essentially equivalent to the conformal
method. Equation (\ref{eq:num_ev}), which we obtained with the
conformal method, can also be derived by using the coordinate based
approach just presented and making consistent choices of rescaling and
coordinate transformations. This has been used as a check for the
final equations that the have numerically implemented.

Note that in the coordinate based method we do not need to rescale the
function $\tphi$ with $\Omega$. We can choose to rescale it with any
function that behaves asymptotically like $r$, and read off the
radiation field directly at null infinity. In fact, a common approach
in numerical calculations is to evolve the variable $\phi:=r\tphi$. To
see how this relates to the conformal rescaling of the previous
section, make a simple choice for the conformal factor
$\Omega=1-\rho$.  In terms of the physical coordinate $r$ we have have
\[ r = \frac{\rho}{1-\rho} \quad \Rightarrow \quad \rho = \frac{r}{1+r}
\quad \Rightarrow \quad \Omega=1-\rho=\frac{1}{1+r}.  \] Therefore and
due to the fall-off behavior of $\tphi$, the rescaling
$\tphi/\Omega=(1+r)\tphi$ is asymptotically equivalent to the
rescaling $r\tphi$ in first order.

\subsection{Hyperboloidal compactification of the scalar wave equation 
in Boyer-Lindquist coordinates}
\label{sec:BL}

We have used the above method to transform the scalar wave equation on
a Kerr background written in Boyer-Lindquist coordinates. The
Kerr metric in those coordinates reads 
\[ \tilde{g}_{\textrm{BL}} = -\left(1-
\frac{2mr}{\widetilde{\Sigma}}\right) dt^2 - \frac{4 a m
  r}{\widetilde{\Sigma}}(1-y^2) \,dt\,d\varphi +
\frac{\widetilde{\Sigma}}{\widetilde{\triangle}}\, dr^2 +
\frac{\widetilde{\Sigma}}{1-y^2}\,d
y^2+\tilde{R}^2\,(1-y^2)\,d\varphi^2.\] Instead of writing the
conformal Kerr metric we derive the scalar wave equation on this
background and work directly with that equation. It is common practice
to introduce the tortoise coordinate $r_\ast$ for the numerical
implementation of the wave equation because the Boyer-Lindquist
coordinates do not give a foliation of the event horizon. The tortoise
coordinate is defined by
\[ \frac{dr_\ast}{dr} = \frac{r^2+a^2}{\widetilde{\Delta}}. \]
The scalar wave equation in Boyer-Lindquist coordinates can be written
as
\begin{eqnarray}\label{eq:bl} \partial_t \tphi_t &=&
  \frac{(r^2+a^2)^2}{D^2}\partial_{r_\ast}\tphi_{r_\ast} +
  \frac{\widetilde{\Delta}(1-y^2)}{D^2}\partial_y \tphi_y + \frac{2 r
    \widetilde{\Delta}}{D^2} \tphi_{r_\ast} - \frac{2 y
    \widetilde{\Delta}}{D^2} \tphi_y, \\ \partial_t \tphi_{r_\ast} &=&
  \partial_{r_\ast}\tphi_t, \qquad \partial_t \tphi_y = \partial_y
  \tphi_t, \qquad \partial_t \tphi = \tphi_t, \nonumber
\end{eqnarray}
where we have defined $D^2:=(r^2+a^2)^2-a^2\widetilde{\Delta}(1-y^2)$.
From the above equation we calculate the equation for a rescaled
variable $\phi:=r \, \tphi$ following the prescription given by
Eq.~(\ref{eq:resc}).  Then we introduce a time function $\tau$ as
in Eq.~(\ref{eq:hypal}), such that the level surfaces of $\tau$
approach null infinity. The Kerr metric in Boyer-Lindquist coordinates
approaches a Schwarzschild metric in standard Schwarzschild
coordinates; therefore, we may assume that the asymptotic form of $H$
needs to be chosen as for the Schwarzschild spacetime in standard
Schwarzschild coordinates. Remember, however, that the asymptotic
Schwarzschild metric in null directions in the tortoise coordinate is
of standard Minkowski form. Therefore, we can choose for $H$ the
expression for the standard hyperboloids in Minkowski spacetime
\cite{Zenginoglu:2007jw}. We set \be \label{eq:height_bl} \Hext =
\frac{r_\ast}{\sqrt{r_\ast^2+C_{\textrm{BL}}^2}} =
\frac{\rho}{\sqrt{\rho+\Omega^2 C_{\textrm{BL}}^2}}. \ee For the
second part of the above equation we used $r_\ast = \rho/\Omega$.
Here $C_{\textrm{BL}}$ is a free parameter that determines the
asymptotic extrinsic curvature of the hyperboloidal surfaces as in the
case of Minkowski spacetime
\cite{Zenginoglu:2007it,Zenginoglu:2007jw}. As in the ingoing Kerr
case, the value of $C_{\textrm{BL}}$ needs to be chosen close to the
value of the coordinate location of null infinity, $S$, in order to
prevent the coordinate speeds of outgoing characteristics from
becoming too large.

For the spatial compactification (\ref{eq:compr}) we choose the
function $\Omega$ as in the ingoing Kerr case. Here, $\Omega$
determines the spatial coordinate compactification and does not
acquire the meaning of a conformal factor for the spacetime metric.

The above choices fix the transformations and result in an explicitly
regular set of equations each of which has the form
(\ref{eq:final}). Their final form is rather lengthy but
straightforward to calculate and will therefore not be presented.

\section{Numerics}
As a test of the two approaches to include null infinity in numerical
simulations by spacelike matching "standard" Cauchy hypersurfaces to
hyperboloidal slicings here we implement them on a $2+1$ code solving
a scalar wave equation on a Kerr spacetime. Near the black hole we use
ingoing Kerr and Boyer-Lindquist coordinates.

The conformal method is applied for the hyperboloidal compactification
in the ingoing Kerr case, and the coordinate based approach in the
Boyer-Lindquist case. We note that there is no specific reason for
this choice. Other coordinate choices are possible as well. The
equations we solve numerically are given in Eq.~(\ref{eq:num_ev}) with
respect to the metric (\ref{eq:kerr_ks}) and in Eq.~(\ref{eq:bl}) with
the transformations (\ref{eq:hypal}), (\ref{eq:compify}), and
(\ref{eq:resc}). The asymptotic form of the height functions for
ingoing Kerr and Boyer-Lindquist foliations are (\ref{eq:height_ks})
and (\ref{eq:height_bl}), respectively.

In both cases we prescribe initial data of the form \be
\label{eq:id} \phi(0,\rho,y)
= 0, \qquad \phi_\tau(0,\rho,y) = e^{-(r-r_0)^2/\sigma^2} \,Y_{l0} \,
\ee where $Y_{l0}$ denotes the familiar spherical harmonics of degree
$l$ and order $k=0$. The data are of compact support. We set
$\phi_\tau(0,\rho,y)=0$ for $\rho>\rhoi$. Regarding the parameters in
the initial data we choose $r_0=5$ and $\sigma=2$ with mass $m=1$.

\subsection{The numerical method}\label{sec:nm}

We use method of lines with fourth order Runge-Kutta time
integration. For the angular derivatives we use a pseudospectral
collocation method with Legendre polynomials, as in
\cite{Tiglio:2007jp}, which is both efficient and automatically takes
care of the coordinate singularities at the poles. We use finite
differences in the radial direction. The reason why we do not use
spectral differencing in the radial direction is the nonsmooth
matching (\ref{eq:match_om}) and (\ref{eq:match_H}). We wish to
remark, however, that we could, in principle, apply spectral
differencing by using a multipatch approach in which the matching
boundaries coincide with some of the patch boundaries. We did not
follow this approach only due to the simplicity of the finite
difference method.

In the ingoing Kerr case the inner boundary is a spacelike surface
inside the event horizon so that one can use the excision
technique. The outer boundary is at null infinity. We apply one-sided
finite differencing at both boundaries using the same order operators
at the boundaries as in the interior \cite{Fornberg}.  In the
Boyer-Lindquist case we set the inner boundary very close to the black
hole. We apply the outgoing boundary condition (ingoing with respect
to the black hole) at the inner boundary, which we set at $r_\ast =
-40m$. We use fourth order accurate finite differencing operators in
the Boyer-Lindquist case instead of the eight order operators used in
the ingoing Kerr case, as the outgoing boundary condition is simpler
to apply with respect to fourth order operators.

We add Kreiss-Oliger type artificial dissipation to the evolution
equation for the time derivative of $\phi$ to suppress numerical
high-frequency waves \cite{Kreiss73}. For a $2p-2$ accurate scheme we
choose an operator $(Q)$ of order $2p$ as
\[ Q=\ \epsilon (-1)^p \frac{h^{2p-1}}{2^p}\,D_+^p D_-^p,  \]
where $h$ is the grid size, $D_\pm$ are the forward and backward
finite differencing operators and $\epsilon$ is the dissipation
parameter which we set as $\epsilon=0.03$.

One major source of error is the matching between the inner
coordinates and the coordinates of the hyperboloidal
compactification. The matching is performed using
Eqs.~(\ref{eq:match_om}) and (\ref{eq:match_H}). We found that the
matching along $\rhoi$ introduces large numerical errors in the
solution that converge with the order of the finite differencing
method. The fact that the numerical errors are large near $\rhoi$ is
to be expected because the matching is not analytic at $\rhoi$ whereas
it is analytic at $\rhoo$. In our simulations we set $\rhoi$ to be at
around $20 m$ in both cases. This is a point where improvement may
lead to a more efficient code. The outer matching introduces no
problems.

We put the numerical outer boundary at $\rho=S=100$ in both cases. The
numerical outer boundary corresponds to future null infinity. Its
coordinate location effects the coordinate speed of characteristics as
discussed before. Experiments with different coordinate locations of
future null infinity delivered qualitatively similar results. An
important point with respect to the coordinate location of the outer
boundary is related to the choice of parameters that determine the
hyperboloidal foliation. The free parameters $C_{\textrm{iK}}$ and
$C_{\textrm{BL}}$ need to be both of the order of $S$ as discussed in
Secs.~\ref{sec:iK} and \ref{sec:BL}.

\begin{figure}[ht]
  \psfrag{t}{$\tau$} \psfrag{Q}{$Q$}
  \includegraphics[height=0.17\textheight]{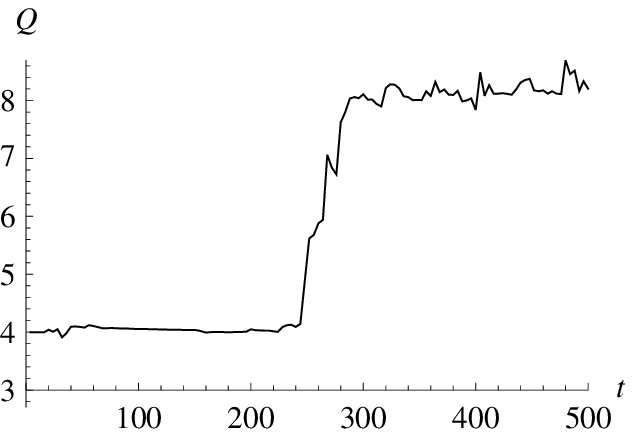}
  \hspace{1cm}
  \includegraphics[height=0.17\textheight]{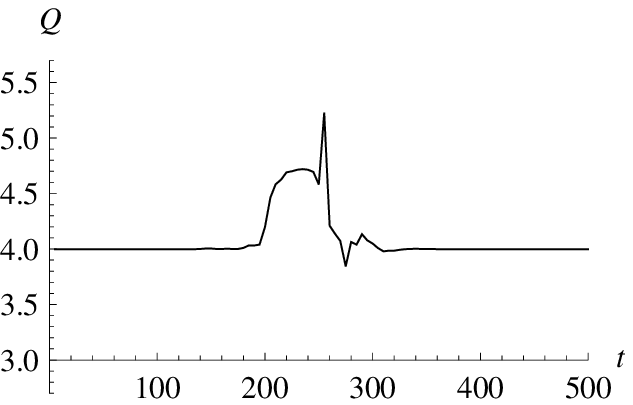}
  \caption{Convergence in the L2 norm in the radial direction for
    ingoing Kerr (left plot) and Boyer-Lindquist (right plot)
    coordinates near the black hole. The convergence factor $Q$ is
    calculated by \mbox{$Q=\log_2\frac{ \|
        \phi^{low}-\phi^{med}\|_{L_2}}{\|
        \phi^{med}-\phi^{high}\|_{L_2}} $} where the norm is taken
    over the whole grid at a given time. We see that in the ingoing
    Kerr case the convergence factor attains its value 8 after a
    transient phase because numerical errors in the time integration
    dominate as observed in \cite{Zenginoglu:2008uc}. In the
    Boyer-Lindquist case the convergence factor is 4. The
    irregularities in the convergence factors are due to
    matching. \label{fig:con}}
\end{figure}

The convergence properties of our codes can be seen in
Fig.~\ref{fig:con}. For these plots, a three level convergence
analysis has been performed for an $l=0$ initial data using 1600, 3200,
and 6400 grid cells in the radial direction and 4 points in the
angular direction where spectral differencing was applied. We used
quadruple precision for the runs because the numerical error was below
machine accuracy at late times when double precision was used. The
time step for the ingoing Kerr case was $dt=0.05$ while for the
Boyer-Lindquist case it was $dt=0.1$. The reason why we can choose a
larger time step in the Boyer-Lindquist case is because the coordinate
speeds of characteristics are lower.

We see that, in the ingoing Kerr case, the convergence factor settles
down at around 8 after an initial transient phase in which it starts
at 4, the order of the time integration. The reason is that in this
initial phase the error is dominated by the time integration. This is
the same behavior as in \cite{Zenginoglu:2008uc}. To get a convergence
factor of 8 from the beginning a much lower time step needs to be
chosen which is not practical for late-time tail calculations
\cite{Zenginoglu:2008uc}. In the Boyer-Lindquist case where both the
time integration and the spatial discretization is of fourth order the
convergence factor is 4. The irregularities in the plots are due to
matching. When a wave package passes through the matching boundary
$\rhoi$, the error increases significantly but in a convergent
fashion.

\begin{figure}[ht]
  \center
  \psfrag{logphi}{$\log|\phi|$}  \psfrag{t}{$\tau$}
  \includegraphics[height=0.17\textheight]{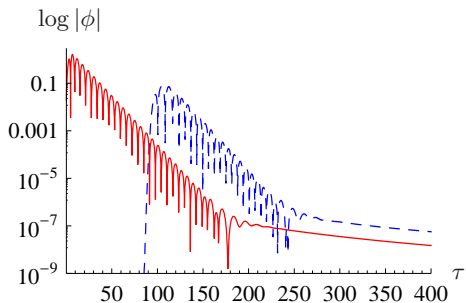}\hspace{1cm}
  \caption{Quasinormal mode ringing of Kerr spacetime with $m=1$ and
    $a=0.5$ excited by a Gaussian initial data with $l=2$ mode as
    measured by an observer located along $r=3m$ (the solid red curve)
    and $\scri^+$ (the dashed blue curve) in ingoing Kerr
    coordinates. \label{fig:qnm}}
\end{figure}
As a next test we calculate a fundamental quasinormal mode
frequency. The quasinormal mode ringing for initial data with $l=2$
can be seen in Fig.~\ref{fig:qnm}. We plot here the ringing as
measured by an observer at $r=3m$ and at $\scri^+$ in ingoing Kerr
coordinates. We observe that the duration of ringing along $\scri^+$
is shorter. Therefore we use the signal measured closer to the black
hole for the fitting.  The frequencies are obtained by fitting the
signal to the formula \be \phi(\tau) = A \,e^{-\omega_2 \tau}
\sin(\omega_1 \tau + \varphi).\ee Here, $\omega_1$ and $\omega_2$ are
the mode frequencies, $A$ is the amplitude and $\varphi$ is the phase
of the wave signal. The fitting is performed using a simple least
squares method on the interval $\tau\in[60m,150m]$. The error in
frequency is rather dominated by the fit than by numerics.  We find
$\omega_1 = 0.491971$ and $\omega_2=0.094653$. These numerical values
are very close to those obtained by using Leaver's continued fraction
method \cite{Leaver85, Leaver:1986gd}, which read $\omega_1 =0.49196$
and $\omega_2=0.09463$. The latter have also been confirmed
numerically by other codes \cite{Krivan97a, Dorband:2006gg}. We obtain
a similar result with the code based on Boyer-Lindquist coordinates.

Quasinormal mode frequencies are the same for all
observers. Including null infinity in the computational domain does
not give us a different picture. The tail decay rates, however, vary
from observer to observer as discussed in the next subsection.

\subsection{Tail decay rates}

In contrast to the quasinormal mode frequencies, the asymptotic tail
decay rates at a finite distance and at null infinity are
different. Also, before the asymptotic decay rates are obtained, the
signals differ substantially from each other. The signal that is
relevant for an idealized observer is the one measured at null
infinity. This is because realistic sources of gravitational radiation
are at astronomical distances from us. Our detectors are better
modeled by observers at future null infinity than by observers at
finite distances in numerical computations \cite{Frauendiener:1998yi,
  Purrer:2004nq, Zenginoglu:2008wc}.

\begin{figure}[ht]
  \psfrag{p}{$p$}\psfrag{t}{$\tau$}
  \includegraphics[height=0.17\textheight]{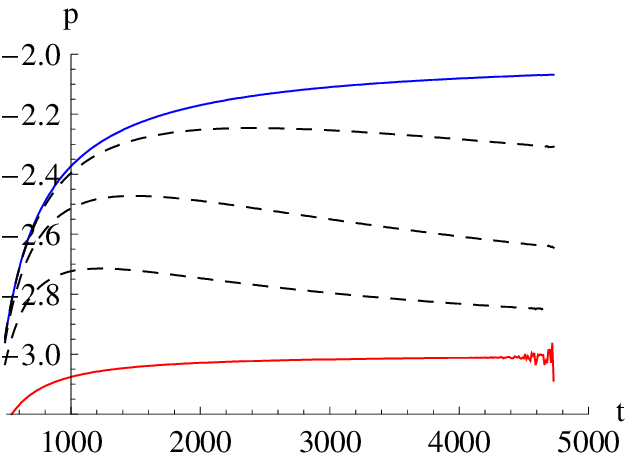}\hspace{1cm}
  \includegraphics[height=0.17\textheight]{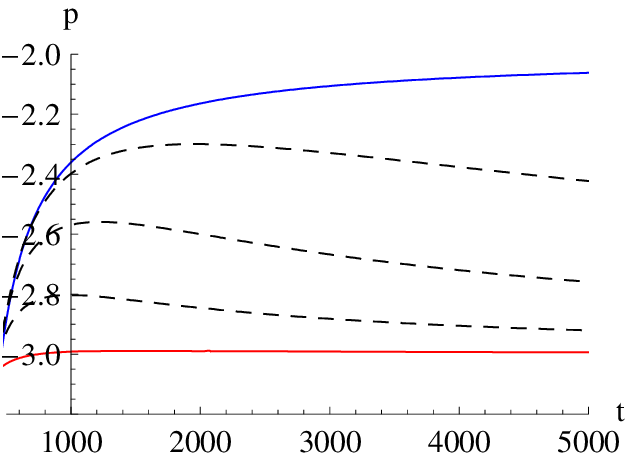}
    \caption{Tail decay rates for initial data with a pure $l=0$ mode
      in ingoing Kerr (left) and Boyer-Lindquist (right)
      coordinates. The solid blue curves are measured by observers
      along $\scri^+$. The solid red curves are measured by observers
      close to the black hole at $r=10$ in units where mass $m=1$. The
      dashed black curves correspond to observers at $r=\{5000, 1000,
      500\}$ from top to bottom. \label{fig:l0}}
\end{figure}

The transition between timelike and null infinity in the decay rates
for a monopole perturbation can be seen in Fig.~\ref{fig:l0}. In
that figure, the local power index has been plotted for various
observers ordered by distance ranging from close to the black hole up
to and including future null infinity. We define the local power index
as \be\label{eq:exponent} p_{\rho,y}(\tau) = \frac{d\ln
  |\phi(\tau,\rho, y)|}{d\ln \tau}.  \ee 
We calculate the time derivatives by using finite differencing from
numerical data. The function $p_{\rho,y}(\tau)$ becomes the exponent
of the polynomial decay of the solution asymptotically in time. We did
not observe a significant variation of the index with respect to the
angular location $y$ of the observer.

\begin{figure}[ht]
  \psfrag{p}{$p$}\psfrag{t}{$\tau$}
  \includegraphics[height=0.17\textheight]{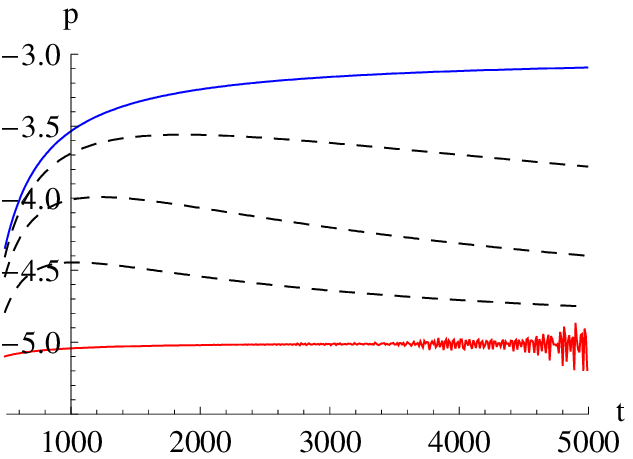}\hspace{1cm}
  \includegraphics[height=0.17\textheight]{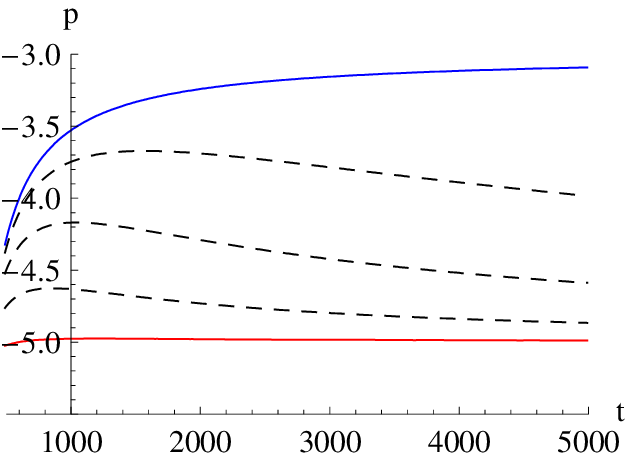}
  \caption{Tail decay rates for an $l=1$ mode. Observers are located
    as in Fig.~\ref{fig:l0}. \label{fig:l1}}
\end{figure}

We see that the monopole perturbation decays with rate $-3$ near
timelike infinity, while the rate at null infinity is $-2$. The decay
rate for the $l=1$ mode is $-5$ near the black hole and $-3$ near null
infinity, as can be seen in Fig.~\ref{fig:l1}. This behavior is the
same as in Schwarzschild spacetime and is in accordance with the
prediction of Table II in \cite{Hod:1999rx}. The decay rate for $l=2$
is as for $l=0$ and the rate for $l=3$ is as for $l=1$ due to mode
coupling in accordance with earlier predictions and calculations
\cite{Hod:1999rx, Burko:2007ju, Gleiser:2007ti, Tiglio:2007jp}.  For
the $l=4$ mode the decay rate near timelike infinity is $-5$ but along
null infinity it is $-4$. This is also in accordance with Hod's
prediction (see Table I in \cite{Hod:1999rx}).

We see that the calculation in ingoing Kerr coordinates near the black
hole is not as accurate as in Boyer-Lindquist coordinates. This is due
to the smaller time step in the ingoing Kerr case as mentioned in
Sec.~\ref{sec:nm}.

We mention that from a physical point of view, in an axisymmetric
spacetime the concept of a pure $l$-mode initial data does not have a
geometrically invariant meaning. In this case, only the monopole
perturbations can be defined invariantly. This has led to some
confusion about decay rates for Kerr spacetime. In our studies the
asymptotic decay rates in both coordinate systems coincide for any
$l$-mode initial data as in (\ref{eq:id}) due to the fact that we are
using the same type of initial data in coordinates that are related in
a way which leaves the notion of an $l$-mode invariant as explained in
\cite{Burko:2007ju}. Here, we found that the asymptotic rates are the
same for such initial data in ingoing Kerr and Boyer-Lindquist
coordinates also at null infinity.

\begin{figure}[ht]
  \psfrag{p}{$p$}\psfrag{t}{$\tau$}
  \includegraphics[height=0.17\textheight]{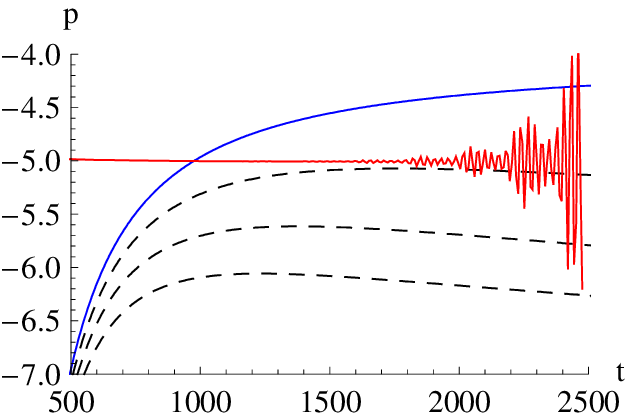}\hspace{1cm}
  \includegraphics[height=0.17\textheight]{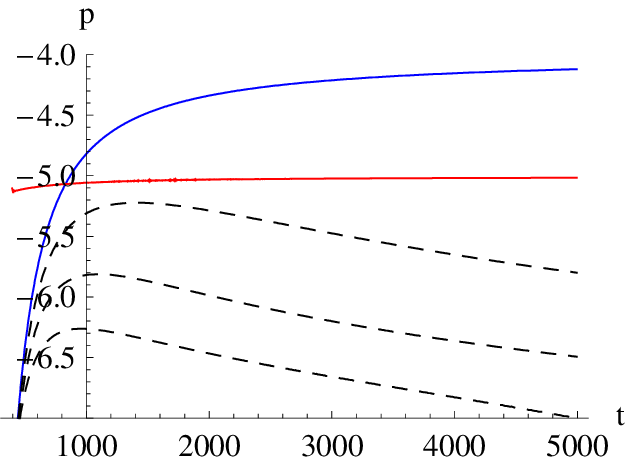}
  \caption{Tail decay rates for an $l=4$ mode. Observers are located
    as in Fig.~\ref{fig:l0}. \label{fig:l4}}
\end{figure}

Predictions of late-time decay rates are valid asymptotically. In our
numerical simulations, we also observe the intermediate behavior of
the local power index. It is well known that for distant observers the
limiting rate is attained later in time \cite{Karkowski:2003et}. One
observes that the decay rate for distant observers is between the rate
for close observers and the rate along null infinity
\cite{Purrer:2004nq,Zenginoglu:2008wc}. This is also seen for the
lower $l$-modes in Kerr spacetime (Figures \ref{fig:l0} and
\ref{fig:l1}). For the $l=4$ mode, however, we observe that the decay
for distant observers is faster than for observers very close to the
black hole (Fig.~\ref{fig:l4}). We do not have analytical evidence
explaining this behavior. It may be suggested that it is due to the
anomalous nature of $l\geq 4$ modes related to mode coupling in Kerr
spacetime \cite{Hod:1999rx, Burko:2007ju, Gleiser:2007ti,
  Tiglio:2007jp}. We speculate that for distant observers the
Schwarzschild decay rates may still have some validity. In
Schwarzschild spacetime, a decay rate of $-4$ at null infinity would
lead to a decay rate of $-7$ along timelike surfaces. This rate seems
to be approached by distant observers in Fig.~\ref{fig:l4}. We
expect that the rate for close by observers will be attained by
distant observers much later in time. We wish to stress, however, that
this is a speculative attempt at explaining what we
observe. Analytical calculations need to be performed to understand
the intermediate behavior of the local power index.

\section{Discussion}

We have presented the first numerical evolution scheme for scalar
fields on the background of a rotating black hole that includes future
null infinity.  Our method is based on the construction of
hyperboloidal surfaces in Kerr spacetime presented in
\cite{Zenginoglu:2007jw}. While previous studies of characteristic
foliations do allow, in principle, compactification at null infinity
on a Kerr spacetime \cite{Venter:2005cs, FletcherLun, Bai:2007rs,
  Pretorius:1998sf}, they did not culminate in numerical
applications. In this paper we have presented and implemented two
methods based on the hyperboloidal compactification technique that
include null infinity in the computational domain. Our methods are
very flexible and allow us to keep any preferred choice of coordinates
near the black hole.

We confirmed well-known results concerning quasinormal mode
frequencies and tail decay rates measured by observers close to the
black hole. In addition, we presented decay rates at and near null
infinity confirming earlier predictions \cite{Hod:1999rx} for the
first time. Our method allows us to study the transition between
timelike and null infinity with respect to the decaying signal.

We expect that our methods can be applied to discuss physically more
interesting evolution equations so that predictions concerning the
behavior of electromagnetic and gravitational perturbations at future
null infinity can be studied as well \cite{Hod:1999ci,
  hod-2000-61}. The coordinate based approach of Sec.~III may be
easier to apply to problems where the conformal transformation
behavior is not known. In numerical calculations the equations of
interest are given anyway in a specific coordinate system. Once the
asymptotic form of the height function is known, the coordinate based
approach can be applied easily to transform the equations to include
null infinity in the computational domain.

In numerical applications of the hyperboloidal approach, one usually
finds that the hyperboloidal method is very efficient in calculating
the outgoing radiation accurately
\cite{Zenginoglu:2008wc,Zenginoglu:2008uc}. In our study of spacelike
matching, however, we find that the presence of a matching domain is a
significant source of numerical error. There are many possibilities to
solve this problem. One option is to devise better matching
techniques, such as matching at a multipatch boundary. This approach
will still have to deal with the change of characteristic speeds
across the matching domain. The other option would be to construct an
analytic, horizon penetrating hyperboloidal foliation which renders
the introduction of matching domains unnecessary. It would be
interesting to study hyperboloidal foliations of Kerr spacetime
analytically or numerically using methods that proved useful in the
case of Schwarzschild spacetime \cite{Malec:2003dq, Malec:2009hg,
  Ohme:2009gn}.

This work suggests that once the spherically symmetric case has been
mastered, extension to less symmetries is more or less straightforward
in the hyperboloidal approach. There are two reasons for this. First,
hyperboloidal foliations can be made to coincide with any known
foliation near the black hole \cite{Zenginoglu:2007jw}. This allows us
to use methods that have been applied successfully in numerical
relativity in an inner domain where the fields are strong. Second,
asymptotically flat spacetimes become flat, and therefore spherically
symmetric in the asymptotic domain. As a consequence the spherically
symmetric case already gives important clues on how to deal with the
asymptotic region in the hyperboloidal approach.

The observations presented in this paper may be useful in the fully
nonlinear case, once the hyperboloidal evolution problem with the
Einstein equations is solved. Specifically, the numerical discussion
of having null infinity at an arbitrary but fixed spatial coordinate
location and setting the conformal factor to unity in the interior
while allowing it to vanish in the asymptotic domain in a certain way
can be expected to be relevant in the nonlinear case.

\acknowledgments

This work was supported in part by the NSF Grant PHY0801213 to the
University of Maryland. We thank F.~Herrmann and T.~Jacobson for
discussions.  
\bibliography{references}
\end{document}